\documentclass[12pt]{article}

\title{
Contest Architecture with Public Disclosures\thanks{
  I would like to thank  
  Federico Boffa,
  Juan Carlos Carbajal,
  Dino Gerardi,
  Marit Hinnosaar,
  Ignacio Monz\'{o}n,
  and Emil Temnyalov 
  for their comments and suggestions.
}}
\date{\today\thanks{
The latest version: \href{https://toomas.hinnosaar.net/contest_architecture.pdf}{\url{toomas.hinnosaar.net/contest_architecture.pdf}.}}}

\author{
Toomas Hinnosaar%
\thanks{Collegio Carlo Alberto, \href{mailto:toomas@hinnosaar.net}{\url{toomas@hinnosaar.net}}. %
}%
}
\usepackage[utf8]{inputenc}
\usepackage[T1]{fontenc}
\usepackage{lmodern}
\usepackage{natbib}
\usepackage{anysize}
\usepackage{subfig}
\usepackage{hyperref}
\usepackage{pgf,tikz}
\usepackage{amssymb,amsmath,amsthm}
\usepackage{thmtools}
\usepackage{multirow}
\usepackage{appendix}
\usepackage{cleveref}
\usepackage{color}

\declaretheorem[name=Proposition]{proposition}
\declaretheorem[name=Theorem]{theorem}

\declaretheorem[name=Definition]{definition}
\declaretheorem[name=Assumption]{assumption}
\declaretheorem[name=Example]{example}

\crefname{figure}{figure}{figures}
\crefname{equation}{equation}{equations}

\newcommand{\dd}[3][]{\frac{d^{#1} #2}{d {#3}^{#1}}}

\newcommand{\bn}{\mathbf{n}}
\newcommand{\bx}{\mathbf{x}}
\newcommand{\hn}{\widehat{n}}
\newcommand{\hbn}{\mathbf{\hn}}
\newcommand{\hX}{\widehat{X}}
\newcommand{\hx}{\widehat{x}}
\newcommand{\hbx}{\boldsymbol{\widehat{x}}}
\newcommand{\oX}{\overline{X}}
\newcommand{\prt}{\mathcal{I}}

\newcommand{\bprt}{\boldsymbol{\mathcal{I}}}
\newcommand{\hbprt}{\boldsymbol{\widehat{\mathcal{I}}}}

\newcommand{\players}{\mathcal{N}}
\newcommand{\bS}{\mathbf{S}}

\newcommand{\uX}{\underline{X}}

\newcommand{\lowx}{\min_i\{x_i^*\}}
\newcommand{\highx}{\max_i\{x_i^*\}}
\newcommand{\lowu}{\min_i\{u_i(\bx^*)\}}
\newcommand{\highu}{\max_i\{u_i(\bx^*)\}}

\begin{document}
\maketitle
\begin{abstract}
  I study optimal disclosure policies in sequential contests. A contest designer chooses at which periods to publicly disclose the efforts of previous contestants. I provide results for a wide range of possible objectives for the contest designer. While different objectives involve different trade-offs, I show that under many circumstances the optimal contest is one of the three basic contest structures widely studied in the literature: simultaneous, first-mover, or sequential contest.
\end{abstract}

\emph{JEL}: C72, C73, D72, D82

\emph{Keywords}: contests, sequential games, contest design, rent-seeking, R\&D, advertising

\section{Introduction} \label{S:intro}

In this paper, I study sequential contests and contest-like economic interactions, where the payoffs increase with participants' own efforts and decrease with the total effort. For example, lobbyists exert efforts to influence politicians towards desired outcomes. The key regulatory decision is transparency---how much and what kind of information should be collected and disclosed to limit the rent-seeking efforts? A fully transparent lobbying disclosure rule would lead to sequential effort choices, whereas a non-transparent policy leads to independent choices of lobbying efforts.\footnote{Different countries have adopted different transparency rules regulating lobbying. For example, in the United States lobbying efforts are all recorded and reported quarterly (Lobbying Disclosure Act, 1995; Honest Leadership and Open Government Act, 2007), whereas in the European Union reporting is arranged on a more voluntary basis and on a yearly frequency (European Transparency Initiative, 2005).} Similarly, firms entering oligopolistic markets invest to increase capacity and the market price decreases with the total capacity. Again, transparency plays a crucial role in these investment decisions and it is natural to ask whether the investments should be publicly observable or not. Many economic interactions satisfy the assumptions of the model discussed here, including oligopolies, public goods provision, rent-seeking, research and development, advertising, and sports.

The objective of the contest designer depends on the specific economic problem. Two standard goals discussed in the literature are maximizing the total effort (such as in research and development contests) or minimizing the total effort (as in rent-seeking contests). However, other objectives are more natural in other situations. For example, a school's objective is not to maximize the average effort, but rather to make sure that all students learn. In fact, the education reform in the US was named the No Child Left Behind Act of 2001. Similarly, in firms where efforts are complementary, it is often important to incentivize the lowest-performing employees, referred to in the business terminology as to as the bottleneck or the weakest link. In welfare economics, there are two standard assumptions. The utilitarian social planner maximizes the total welfare of the participants, whereas the Rawlsian social planner maximizes the lowest utility.

Different objectives lead to different trade-offs and it is natural that the optimal contests in various economic interactions may differ. However, the literature has so far almost exclusively focused on three standard types of contests. First, \emph{simultaneous contests}, where the players make their choices independently. This is the least informative form of contest, where no efforts are disclosed. It is the most common type of contest considered by the literature, starting with \cite{cournot_recherches_1838} in oligopoly theory and \cite{tullock_welfare_1967,tullock_social_1974} in contest theory. 
The second type of contest that has been extensively studied is the \emph{first-mover} contest, where a single first-mover chooses the effort first and the rest of the players move simultaneously in the second period. In oligopoly theory it was first analyzed by \cite{von_stackelberg_marktform_1934} and in contest theory by \cite{dixit_strategic_1987}.\footnote{For literature reviews on dynamic contests, see \cite{konrad_strategy_2009} and \cite{vojnovic_contest_2015}.} The third type of contest that has been considered is the \emph{sequential contest}, where all effort choices are public. Then each player observes all the previous effort choices. Simultaneous contests have been studied by \cite{robson_stackelberg_1990} in the case of large oligopolies, \cite{glazer_sequential_2000}, \cite{hinnosaar_optimal_2018}, and \cite{kahana_sequential_2018} in the case of sequential Tullock contests, and \cite{hinnosaar_optimal_2018} with more general payoff functions.

The goal of this paper is to understand which types of contests are optimal under different circumstances. On one hand, the paper aims to provide a menu of optimality results that researchers and practitioners could build upon. Perhaps even more importantly, finding contest structures that are optimal under many different sets of assumptions may explain why these types of contests are often used.

The main difficulty in studying sequential contests is tractability. Sequential games with non-quadratic payoffs are typically difficult to solve because best-response functions are non-linear and therefore the standard backward-induction approach leads to increasingly complex expressions. This problem is even more pronounced in contest design, where the goal is to compare all possible contests. In this paper, I build on recent progress in aggregative games \citep{jensen_aggregative_2010,martimort_representing_2012,acemoglu_aggregate_2013} and sequential contests \citep{hinnosaar_optimal_2018,hinnosaar_stackelberg_2019} to overcome the tractability issues. 

The main finding of the paper is that under many circumstances the optimal contest is one of the three basic contest structures widely studied in the literature. In particular, I consider minimizing and maximizing eight different objectives: total effort, total welfare, lowest effort, lowest payoff, highest effort, highest payoff, effort inequality, and payoff inequality. I show that nine out of the sixteen maximization problems are solved either by the simultaneous contest or by the sequential contests. With the four out of the remaining five objectives I show that when the number of players is sufficiently large, they are maximized by one of the three basic contest structures mentioned above. I also provide conditions under which the same conclusions hold when the number of players is small. Finally, there is a single exception: minimizing the highest payoff, which typically requires a non-standard contest.\footnote{I provide conditions under which it is maximized by a contest where the first two players are moving simultaneously and all the remaining players are arranged sequentially.}

The paper contributes to both the contest design and the information design literatures. Earlier papers on contest design have focused on contests with private information and have studied how to arrange contests into subcontests and which prizes to offer to maximize either total effort or highest effort \citep{glazer_optimal_1988,taylor_digging_1995,che_optimal_2003,moldovanu_optimal_2001,moldovanu_contest_2006,olszewski_large_2016,bimpikis_designing_2019}.
In this paper, I study contest design with full information. The designer may have different objectives and can choose to disclose less or more information about the players' efforts.
There is growing literature on information design \citep{kamenica_bayesian_2011,bergemann_information_2019}, which has mostly focused on revealing information about the state of the world or private information. This paper is more in line with recent papers that have also studied disclosure of the actions of other players \citep{doval_sequential_2019,ely_moving_2019,gallice_co-operation_2019}.

The paper is organized as follows. \Cref{S:model} describes the model. \Cref{S:results} provides all results, describing the contests that minimize and maximize eight different objective functions. \Cref{S:discussion} summarizes and concludes. Proofs are in the appendix.

\section{Model} \label{S:model}

There is a finite number $n$ players, $\players = \{1,\dots,n\}$. Players arrive sequentially and each player $i \in \players$ chooses effort level $x_i \geq 0$ at arrival. For a given profile of efforts $\bx = (x_1,\dots,x_n)$, the payoff of player $i$ is $u_i(\bx) = x_i h(X)$, where $X = \sum_{i=1}^n x_i$ denotes the total effort and $h(X)$ is the marginal benefit of effort, which is common to all players (I discuss its properties below).

The contest designer chooses points of time at which to publicly disclose the cumulative effort of all players that have already arrived. The disclosure rule is chosen before effort choices and is commonly known. Each such disclosure rule induces a partition of players $\bprt = \left(\prt_1,\dots,\prt_T\right)$, where $\prt_t$ is the set of players arriving between disclosures number $t-1$ and $t$. I refer to the set of players $\prt_t$ as players arriving in period $t$. 
Then player $i \in \prt_t$ observes the cumulative effort of all players arriving in previous periods, $X_{t-1} = \sum_{s=1}^{t-1} \sum_{j \in \prt_s} x_j$, and makes a choice simultaneously with all players arriving in period $t$.
As all players are identical, the disclosure rule is equivalently defined by a vector $\bn = (n_1,\dots,n_T)$, where $n_t = \# \prt_t$, so that $\sum_{t=1}^T n_t = n$.\footnote{Alternatively, the contest designer's problem can be interpreted as dividing players between periods.}

The marginal benefit function $h(X)$ is a strictly decreasing and continuously differentiable function in $[0,\oX]$, where $\oX > 0$ is the saturation point so that $h(\oX)=0$. Moreover, I impose two technical assumptions, which are discussed in \cref{A:assumptions}. The first assumption guarantees that the best-response functions are well-behaved so that a subgame-perfect Nash equilibrium exists and is unique. The second assumption guarantees that the efforts are higher-order strategic substitutes near equilibrium, i.e.\ higher effort of an earlier-mover discourages efforts by followers both directly and indirectly (through the changed efforts of other players). Both assumptions are satisfied for most applications, including the following three examples.

\begin{example}[Sequential Tullock Contest] 
Players choose efforts to compete for a prize with value $v>0$ and have constant marginal cost $c>0$. The probability of winning is proportional to efforts. The payoffs can be expressed as $u_i(\bx) = \frac{x_i}{X} v - c x_i = x_i h(X)$, where $h(X) = \frac{v}{X}-c$.
\end{example}

\begin{example}[Tragedy of the Commons]
The total amount of resources is $1$ and each player chooses private consumption $x_i \geq 0$. Marginal benefit of consumption is linearly decreasing in the total consumption, i.e.\ $h(X)=1 - X$.\footnote{This example can be interpreted as private provision of public goods, where each player divides endowment $\omega_i$ between private consumption $x_i$ and public good contribution $g_i = \omega_i - x_i$. Suppose that the total endowment is $1$, so that the total quantity of public good is $\sum_{i=1}^n g_i = 1-X$. The payoff is a multiplicative function of both private consumption and public good, for example $u_i(\cdot)= x_i (1-X)$. The marginal benefit function $h(X)$ can be non-linear.}
\end{example}

\begin{example}[Oligopoly with a Non-linear Demand Function]
Suppose that $n$ oligopolists have identical constant marginal costs $c \geq 0$ and produce identical good. Each oligopolist chooses a quantity (capacity) and the inverse demand function is $P(X) = a - b^X$ for some $a>1$ and $b>1$. 
The profit of firm $i$ is $u_i(\bx) = x_i h(X)$, where $h(X)=P(X)-c$.
\end{example}

To compare different contests, I define \emph{informativeness} as a partial order on all $n$-player contests. Intuitively, a contest $\bn$ is equivalently defined by a partition $\bprt$ of players. Each additional disclosure divides the players in one period $\prt_t$ into two groups. Therefore the new contest provides strictly more information to some players than the previous contest, whereas no player receives less information. Formally, I say that the contest $\hbn$ is more informative than the contest $\bn$, if the corresponding partition $\hbprt$ is finer than $\bprt$. 

In particular, the \emph{simultaneous contest} $\bn=(n)$ that does not provide any information about the efforts of other players is less informative than any other contest. In the other extreme, the \emph{sequential contest} $\bn=(1,1,\dots,1)$ discloses efforts after each player and is therefore more informative than any other contest. Finally, the \emph{first-mover contest} $\bn=(1,n-1)$ discloses information after the first player, whereas all other players make their moves simultaneously. The first-mover contest is less informative than any other \emph{single-leader} contest $\bn=(1,n_1,\dots,n_T)$.

\section{Optimal Contests} \label{S:results}


\subsection{Total Effort and Total Welfare} 

The most common objectives considered in the literature are minimizing or maximizing the total equilibrium effort $X^*$. \cite{hinnosaar_optimal_2018} showed that the total effort is strictly increasing in the informativeness of the contest, i.e.\ each additional disclosure increases the total equilibrium effort $X^*$. Therefore the simultaneous contest minimizes and the sequential contest maximizes the total equilibrium effort. 
The intuition for this result comes from the discouragement effect. Efforts in contests are strategic substitutes. Therefore, with each additional public disclosure, each earlier-mover whose effort is now made visible to some additional later-movers has an additional incentive to increase the effort---it discourages these later-movers from exerting high effort. Moreover, in equilibrium this effect must be less than one-to-one, i.e.\ the decrease in efforts by later-movers is smaller than the increase in efforts by earlier-movers. Otherwise, a marginal increase in the effort $x_i$ of an earlier-mover $i$ would decrease the total effort $X$ and therefore increase $i$'s payoff. This would be a profitable deviation and thus violate equilibrium conditions.

Let us now look at total equilibrium welfare. For example, in the case of a normalized Tullock contest the total welfare is $W(\bx^*) = \sum_{i=1}^n u_i(\bx^*) = 1 - X^*$. This expression is decreasing in $X^*$, so maximized by the simultaneous contest and minimized by the sequential contest. The intuition is simple---in a Tullock contest, the total prize is fixed. Additional players and additional disclosures lead to a higher total effort, which is costly and therefore welfare-reducing.
The following proposition shows that these conclusions generalize to arbitrary payoffs and an arbitrary number of players.

\begin{proposition}[Total Effort and Total Welfare] \label{P:totals}
  The total equilibrium effort $X^*$ is minimized by the simultaneous contest and maximized by the sequential contest.
  The total equilibrium welfare $W(\bx^*)=\sum_{i=1}^n u_i(\bx^*)$ is minimized by the sequential contest and maximized by the simultaneous contest.
\end{proposition}

\subsection{Lowest Effort and Lowest Payoff}

We need to first determine who is the player choosing the lowest effort in equilibrium and who gets the lowest payoff. The answer comes from the earlier-mover advantage---equilibrium efforts and payoffs of earlier-movers who are observed by more players are higher than the corresponding efforts and payoffs of later-movers.\footnote{See \cref{A:assumptions} for a formal discussion.} Therefore player $n$ chooses the lowest effort and gets the lowest payoff with any $n$-player contest $\bn$. The reason for this result is again the discouragement effect---earlier players choose higher efforts to discourage later players from exerting effort whenever their efforts are made public. By doing so, earlier players ensure higher and later players get lower payoffs.

The following proposition shows that both the lowest effort and the lowest payoff are minimized and maximized by the same contests as the total welfare---sequential and simultaneous contests respectively. 
The intuition is simple. As discussed above, each additional disclosure leads to higher efforts by earlier-movers and lower efforts by later-movers. Therefore the discouragement effect to the last player $n$ is increased with each disclosure. Moreover, additional disclosures lead to higher total equilibrium effort $X^*$, which reduces the payoff of the last player even further.

\begin{proposition}[Lowest Effort and Lowest Payoff] \label{P:lowest}
  Lowest equilibrium effort $\lowx$ and lowest equilibrium payoff $\lowu$ are minimized by the sequential contest and maximized by the simultaneous contest.
\end{proposition}

\subsection{Highest Effort and Highest Payoff}

\begin{figure}[!ht]%
    \centering
    \subfloat[Highest effort. Green dash-dotted line indicates the single-leader contest with pairwise followers. The slightly lower red dash-dotted line indicates the first-mover contest.
    ]{\includegraphics[trim={0.22cm 0.35cm 0.86cm 0.35cm},clip,width=0.45\linewidth]{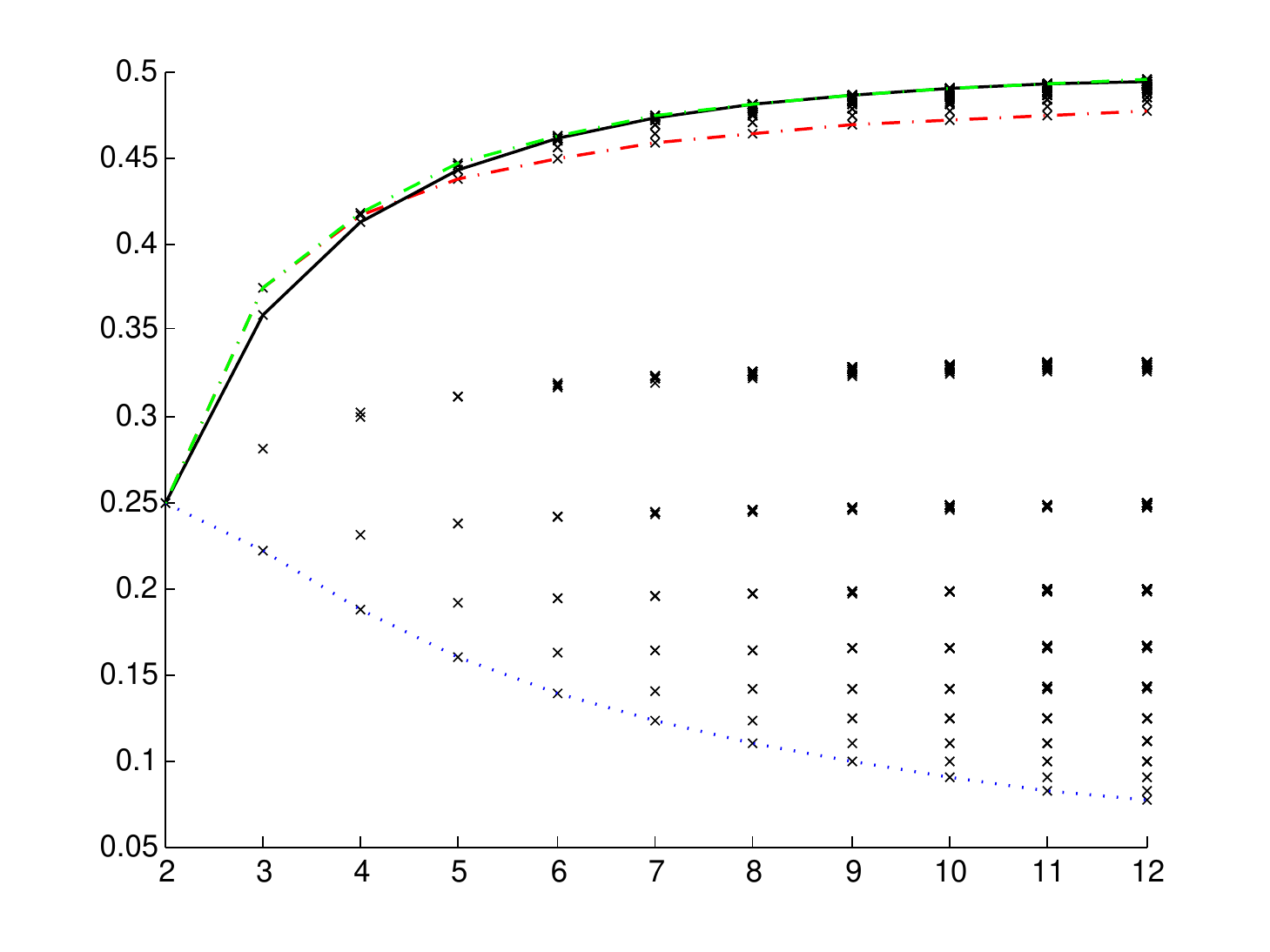} \label{F:plotx1}}
    \qquad
    \subfloat[Highest payoff. Green dash-dotted line indicates contests $\bn=(1,n-1)$ and red long-dashed line indicates contests $(2,1,\dots,1)$.]{\includegraphics[trim={0.22cm 0.35cm 0.86cm 0.35cm},clip,width=0.45\linewidth]{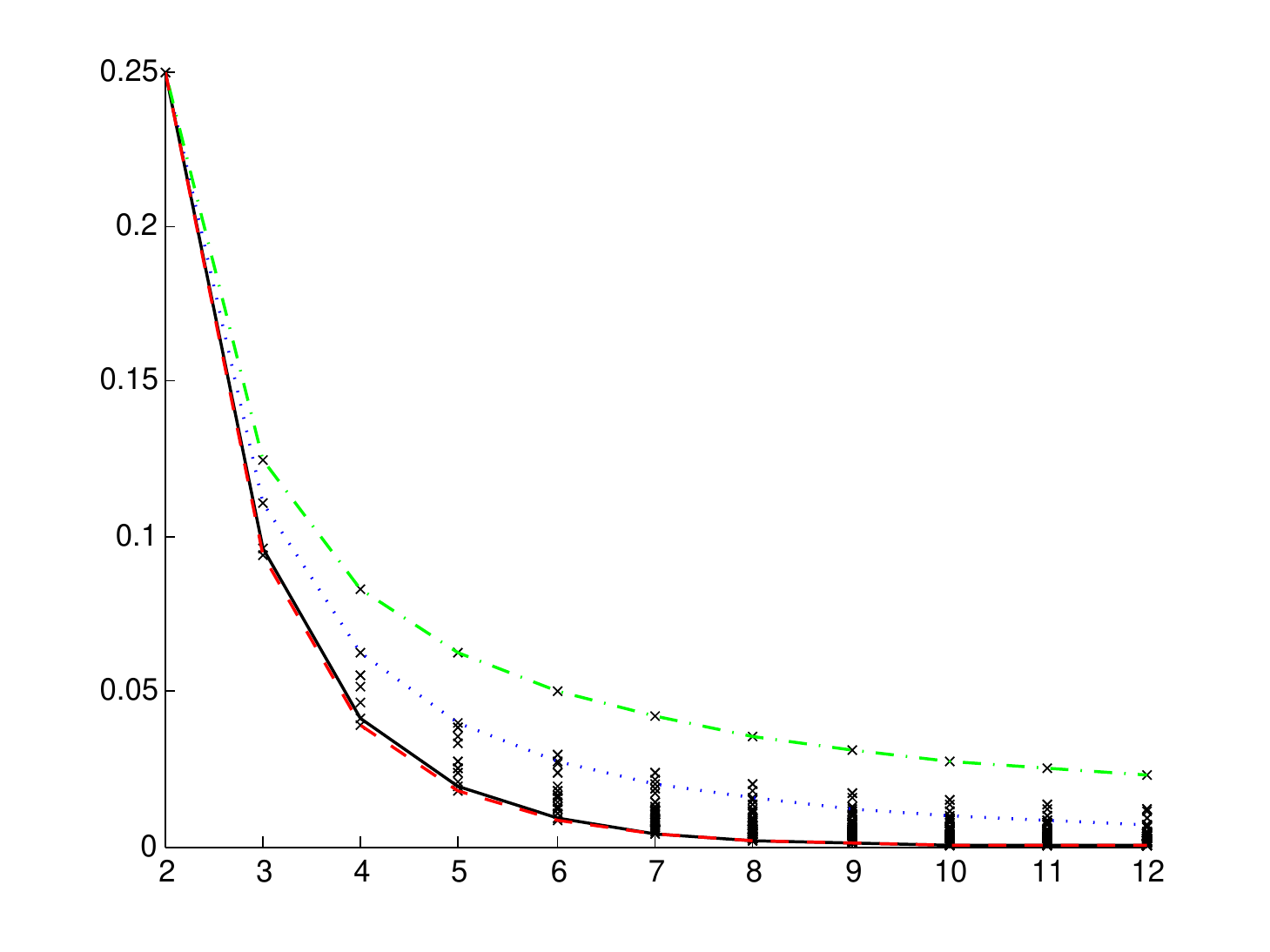} \label{F:plotu1}}
    \caption{Highest effort $\highx$ and highest payoff $\highu$ in all normalized Tullock contests with $2 \leq n \leq 12$ players. The solid black line indicates sequential contests and dashed blue line simultaneous contests.
    \label{F:plotx1u1}}%
\end{figure}

The earlier-mover advantage result shows that the player who exerts the highest effort and gets the highest payoff is the first player.
It is easy to see that the effort of player $1$ is minimized by the simultaneous contest. This is the contest which minimizes the total effort $X^*$ and where all efforts are equal. In all other contests, the total effort is higher and the effort of player $1$ is higher than the average, due to the discouragement effect, which means higher than in the simultaneous contest. Indeed, this is what \cref{F:plotx1} confirms---for each $n$, the highest equilibrium effort is minimized by the simultaneous contest.

Maximization of the highest effort $x_1^*$ is the first objective, where the optimal solution is not always one of the extremes, i.e.\ simultaneous or sequential contest. For example, let us consider three-player Tullock contests with a single leader.\footnote{In numerical examples, I normalize the prize and marginal cost to $1$, so that the payoff of player $i$ is $u_i(\bx)  = \frac{x_i}{X} - x_i$. For figures, I consider $\sum_{n=2}^{12} 2^{n-1} = 4094$ possible contests with $2$ to $12$ players. I use the Matlab code available at \href{https://toomas.hinnosaar.net/contests/}{\url{toomas.hinnosaar.net/contests/}}.
} There are two such contests. The first-mover contest $\bn=(1,2)$ has a total equilibrium effort $X^* = 0.75$ and the highest effort $x_1^* =0.375$. Adding a disclosure after player $2$ makes the contest sequential, i.e.\ $\hbn=(1,1,1)$. As argued above, this raises the total effort to $\hX^* \approx 0.7887$. This reduces the marginal benefit of effort for all players. However, it also makes the discouragement effect of player $1$ stronger through indirect influence. Note that players $2$ and $3$ already observed the effort of player $1$, so the direct discouragement effect is unchanged. However, in the sequential contest player $3$ observes player $2$, who is influenced by player $1$. In the case of Tullock contest payoffs the first effect dominates and the leader's equilibrium effort reduces to $\hx_1^* \approx 0.3591 < x_1^*$.
Numerical calculations summarized by \cref{F:plotx1} show that the contest that maximizes the  highest equilibrium effort in a Tullock contest with $2< n \leq 12$ players is always one with a single leader and followers that are arranged pairwise, i.e.\ $\bn=(1,2,\dots,2)$ when $n$ is odd or $\bn =(1,2,\dots,2,1)$ when $n$ is even.\footnote{The location of the second single-player period does not affect $x_1^*$.} Arranging the followers pairwise leads to a strictly higher effort by the first player than in any other contest, including the sequential contest.

The observation that there should be a single first-mover is general and intuitive. Making first-movers' effort observable to all players increases total effort only because of increased effort by the first player, so there is no trade-off.
As discussed above, each additional disclosure has two opposite effects and the effect on highest effort depends on the weights to direct and indirect influences. For example, with exponential function $h(X)=\frac{1}{\log 2} \left[ 2^{-X}-2^{-1} \right]$ the first-mover contest $\bn=(1,2)$ gives highest effort $x_1^* \approx 0.3698$, whereas the sequential contest $\hbn=(1,1,1)$ gives strictly larger highest effort $\hx_1^* \approx 0.3714 > x_1^*$.
The literature on sequential oligopolies (starting from \citep{daughety_beneficial_1990}) has found that if the demand function is linear, then Stackelberg leaders behave as if there are no followers, i.e.\ $x_1^*$ would be independent on the number and arrangement of followers. \cite{hinnosaar_stackelberg_2019} provides the most general formulation of this Stackelberg independence result, but also shows that this result holds only when $h(X)$ function is linear. 
Therefore the formal statement, in this case, provides only a qualitative result, showing that $n_1=1$. Indeed, as any single-leader is optimal in the case of linear $h(X)$, the result cannot exclude any such contest from being optimal.

Finally, in the case when $n$ is large enough, we can say more regardless of the exact details of the payoff function. Namely, in this case, $X^*$ is close to its upper bound and therefore assuming that $h(X)$ function is smooth, it can be closely approximated by a linear function. In particular, this means that the conclusion from linear demand extends here---any large single-leader contest leads to the highest effort that is arbitrarily close to the maximized highest effort. \Cref{F:plotx1} also illustrates that while all single-leader contests are optimal in the limit, for the sequential contest the convergence is very fast, whereas for the first-mover contest the convergence is much slower.\footnote{The same arguments can be used to find the contest that minimizes or maximizes the $i$-th highest effort or payoff when $h(X)$ is linear or $n$ is large enough. \Cref{A:ithlargest} provides these results.}

\begin{proposition}[Highest Effort] \label{P:highestx}
  The highest effort $\highx$ is minimized by simultaneous contest.
  If contest $\bn$ maximizes the highest effort then $n_1 = 1$.
  If $h(X)$ is linear or $n$ is large enough, then any single-leader contest ensures highest effort that is arbitrarily close to maximum.
\end{proposition}

Maximizing highest equilibrium payoff $u_1(\bx^*)$ balances a similar trade-off, but the downward force through reduced marginal benefit $h(X^*)$ is more pronounced. Not only does it push effort down, but it also reduces $h(X^*)$, which directly reduces all payoffs. 
This means that we would naturally expect the optimal contest to have fewer disclosures than the contest that maximizes the highest equilibrium effort.
Indeed, \cref{F:plotu1} shows that among all Tullock contests with up to $12$ players, the first-mover contest $\bn =(1,n-1)$ is always optimal.

The proposition below formalizes this qualitative property---a contest that maximizes the highest equilibrium effort must have a single leader and has to be (weakly) less informative than a contest that maximizes the highest payoff. For example, in the case of linear $h(X)$ function, one maximizer of the highest effort is the first-mover contest $\bn=(1,n-1)$. As all other contests with a single leader are more informative, the immediate conclusion is that the first-mover contest is the unique maximizer of the highest payoff.
In the case of linear $h(X)$ function or a large number of players, we can therefore uniquely determine the optimal contest. It is always $\bn=(1,n-1)$, which confirms the finding from \cref{F:plotu1}.

Minimizing the highest payoff $u_1(\bx^*)$ requires balancing a different trade-off. 
As we saw above, additional disclosures lead to larger highest effort due to a stronger discouragement effect. This would imply that minimizing the highest payoff requires a relatively uninformative contest. On the other hand, additional disclosures lead to a higher total effort, which reduces all payoffs.
Let us compare for example three-player Tullock contests $\bn=(2,1)$ and $\hbn=(1,1,1)$. Contest $\bn$
gives total effort $X^*=0.75$, highest effort $x_1^* \approx 0.2813$ and therefore highest payoff $u_1(\bx^*)\approx 0.0938$. Contest $\hbn$ gives total effort $\hX^*\approx 0.7887>X^*$, but highest effort $\hx_1^* \approx 0.3591$, which is much larger than $x_1^*$ and therefore highest payoff $u_1(\hbx^*)=0.0962 > u_1(\bx^*)$.
\Cref{F:plotu1} shows that in the case of a Tullock contest with up to $12$ players, contest $\bn = (2,1,\dots,1)$ minimizes the highest payoff.

The formal result below provides only one general qualitative property for the contest that minimizes the highest payoff: the number of players in the first period must be weakly higher than in any other period, i.e.\ $n_1 \geq n_t$ for all $t$. If this would not be the case, the designer could rearrange the groups while keeping the total effort unchanged and decreasing the highest effort, which would reduce the highest payoff.\footnote{This comes from the independence of permutations result from \cite{hinnosaar_optimal_2018}, which shows that total equilibrium effort $X^*$ is independent of permutations of vector $\bn=(n_1,\dots,n_T)$.}
Why can't we say more? Other possible manipulations of the contests, such as moving a player from period 1 to period 2 or splitting up the period by an additional disclosure, lead to two opposite effects: on one hand they increase the total effort in the contest, which reduces all efforts as well as the payoffs directly, but on the other hand they increase the influence that the first player has through the discouragement effect. The relative magnitudes of these effects depend on the shape of the function $h(X)$.

In the case of linear $h(X)$ or large contests, the optimal contest is again uniquely determined: it is always $(2,1,\dots,1)$.\footnote{Remark: when $n$ is very large, $h(X^*) \to h(\oX)=0$ and therefore $u_i(\bx^*) \to 0$ for all $i$, which means that technically all contests are approximately optimal. However, as \cref{F:plotu1} illustrates, in some contests this convergence is much faster than in others. By optimality I mean here the contest for which the value of the objective is maximized for large but finite values of $n$.}
This confirms the finding from \cref{F:plotu1}.

\begin{proposition}[Highest Payoff] \label{P:highestu}
  If contest $\bn=(n_1,\dots,n_T)$ minimizes the highest equilibrium payoff $\highu$, then $n_1 \geq n_t$ for all $t$. If $h(X)$ is linear or $n$ is large enough, then the optimal contest is $\bn=(2,1,1,\dots,1)$.
  
  If contest $\bn$ maximizes the highest equilibrium payoff, then $n_1=1$ and $\bn$ is not more informative than any maximizer of the highest effort. If $h(X)$ is linear or $n$ is large enough, then the optimal contest is $\bn=(1,n-1)$.
\end{proposition}

\subsection{Effort Inequality and Payoff Inequality}

In this final part I study equilibrium effort inequality, defined as $\highx-\lowx$, and equilibrium payoff inequality, defined as $\highu-\lowu$. By the earlier-mover advantage result discussed above, these expressions are equivalent to $x_1^*-x_n^*$ and $u_1(\bx^*)-u_n(\bx^*)$ respectively.
It is easy to see that both expressions are minimized by the simultaneous contest since this is the only contest where efforts and payoffs are equal for all players.

\begin{figure}[!ht]%
    \centering
    \subfloat[Effort inequality]{\includegraphics[trim={0.22cm 0.35cm 0.86cm 0.35cm},clip,width=0.45\linewidth]{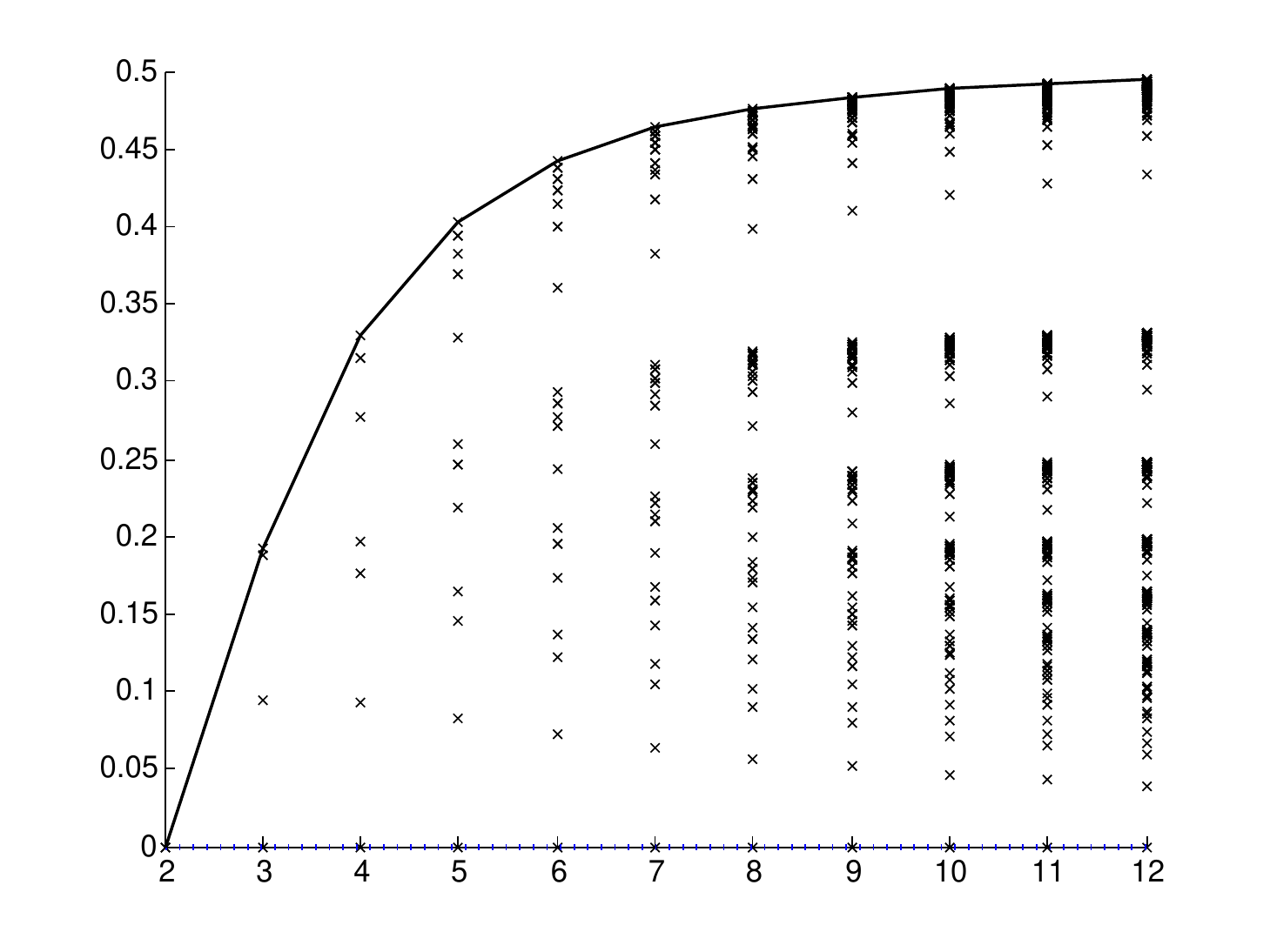} \label{F:plotdx}}
    \qquad
    \subfloat[Payoff inequality. Green dot-dashed line indicates the first-mover contest.]{\includegraphics[trim={0.22cm 0.35cm 0.86cm 0.35cm},clip,width=0.45\linewidth]{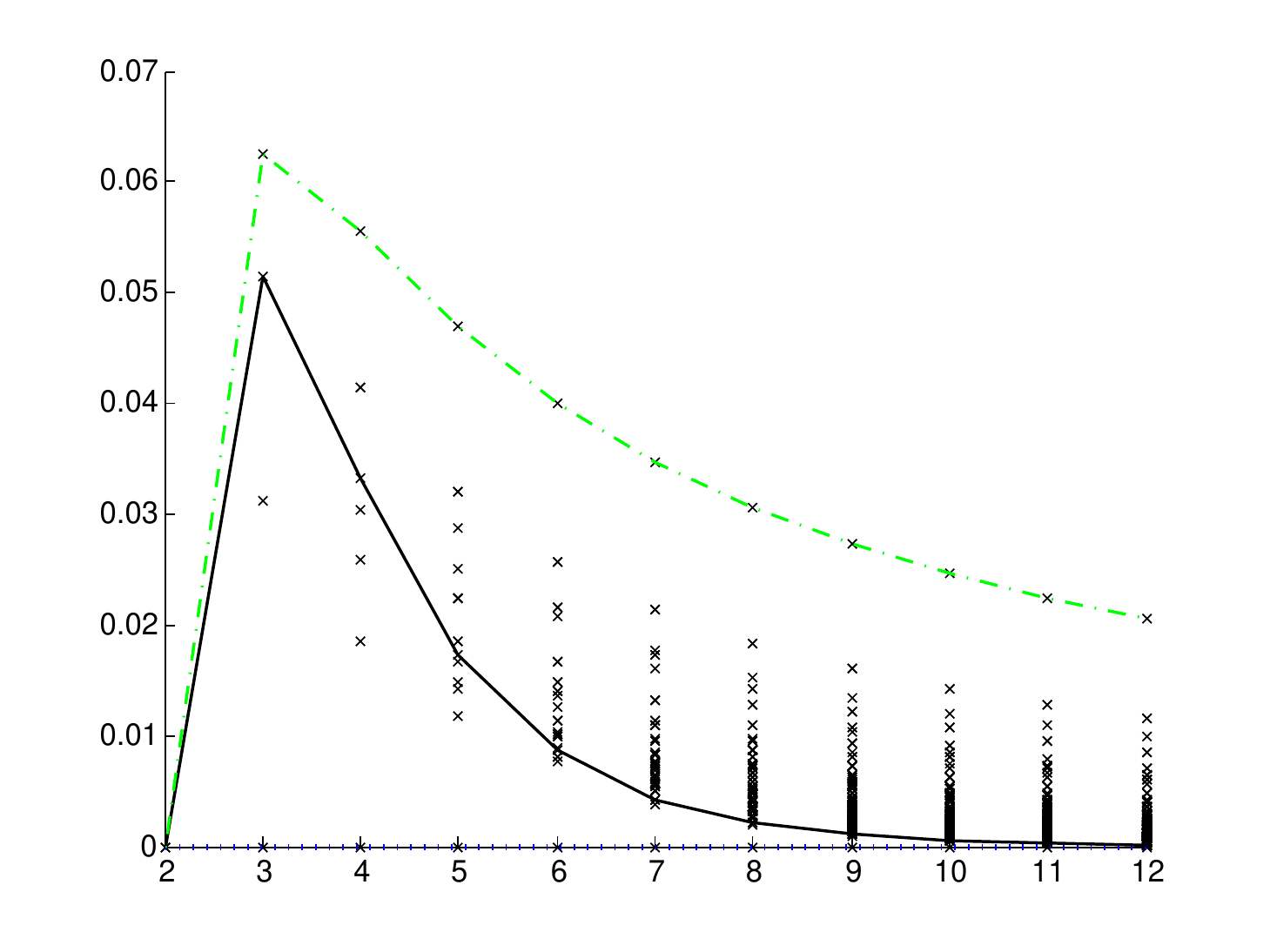} \label{F:plotdu}}
    \caption{Effort inequality $\highx-\lowx$ and payoff inequality $\highu-\lowu$ in all normalized Tullock contests with $2 \leq n \leq 12$ players. The solid black line indicates sequential contests and simultaneous contests lie on the horizontal axis.
    \label{F:plotdxdu}}%
\end{figure}

Maximization of effort inequality and payoff inequality leads to additional trade-offs. Consider the effort inequality first. As seen above, minimizing the lowest effort requires an informative contest. On the other hand, maximizing highest effort may involve pooling some of the later-movers. Therefore the effort inequality is maximized by a contest that is (weakly) more informative than the contest that maximizes the highest effort.
Indeed, as \cref{F:plotdx} shows, in Tullock contests with up to $12$ players the first effect dominates and the optimal contest is sequential. Let us compare again Tullock contests $\bn=(1,2)$ and $\hbn=(1,1,1)$. The first contest leads to maximized highest effort $x_1^*=0.375$, but also relatively high lowest effort $x_3^*=0.1875$. The sequential contest $\hbn$ leads to slightly lower highest effort $\hx_1^*\approx 0.3591$, but the impact to the lowest effort is bigger, $\hx_3^* \approx 0.1667$. Therefore the sequential contest maximizes the effort inequality.

The formal proposition proves these qualitative properties. It shows that the payoff-inequality-maximizing contest must have a single leader and it cannot be less informative than any contest that maximizes the highest effort. In the case when $h(X)$ is linear or $n$ is large, the most informative maximizer of the highest effort is the sequential contest. Therefore, we can immediately conclude that the sequential contest must also be the unique maximizer of the equilibrium effort inequality, which confirms the result from \cref{F:plotdx}.

Finally, maximizing payoff inequality has an additional first-order effect: disclosures increase $X^*$ and thus reduce all payoffs. This effect mechanically reduces payoff inequality as $u_1(\bx^*)-u_1(\bx^*)
=(x_1^*-x_n^*) h(X^*)$.
In the three-player examples above, $x_1^*-x_n^*=0.1875 < \hx_1^*-\hx_3^* \approx 0.1925$. However, $h(X^*) \approx 0.3333 < \hX^* \approx 0.2679$. Therefore this additional effect is larger and $\bn=(1,2)$ is the three-player contest that maximizes payoff inequality.
\Cref{F:plotdu} shows that in the case of Tullock contests, this effect is strong enough so that for all $n \leq 12$, the contest that maximizes payoff inequality is always the first-mover contest $\bn=(1,n-1)$.

The proposition below shows that the contest that maximizes payoff inequality must have a single leader. It is more informative than any maximizer of the highest payoff and less informative than any maximizer of the effort inequality.
When $h(X)$ function is linear or the contest is large, these conditions, unfortunately, are not very informative. All single-leader contests are more informative than the highest payoff maximizer $(1,n-1)$ and all contests are less informative than the effort inequality maximizer, which is the sequential contest.
The proposition shows that we can in this case nevertheless uniquely determine that the optimal contest is the first-mover contest $\bn = (1,n-1)$.\footnote{Again, subject to a remark: when $n$ becomes large then all payoffs converge to zero and therefore any contest approximately maximizes the payoff inequality. However as \cref{F:plotdu} clearly illustrates, $u_1(\bx^*)-u_n(\bx^*)$ in the first-mover contest converges to zero much faster than other contests. The optimal contest here means that it maximizes the payoff inequality with finite but arbitrarily large $n$.}

\begin{proposition}[Effort Inequality and Payoff Inequality] \label{P:inequality}
  Both the equilibrium effort inequality $\highx-\lowx$ and the equilibrium payoff inequality $\highu-\lowu$ are minimized by the simultaneous contest.
  
  If contest $\bn$ maximizes the equilibrium effort inequality, then $n_1=1$ and $\bn$ is not less informative than any maximizer of the highest effort. If $h(X)$ is linear or $n$ is large enough, then $\bn$ is the sequential contest.
  
  Any contest $\bn$ that maximizes the equilibrium payoff inequality must satisfy $n_1=1$. Moreover, it cannot be less informative than any maximizer of the highest payoff and it cannot be more informative than any maximizer of the effort inequality. If $h(X)$ is linear or $n$ is large enough, then $\bn$ is the first-mover contest.
\end{proposition}

\section{Conclusions} \label{S:discussion}

In this paper, I studied contest architecture with public disclosures. Additional disclosures induce the discouragement effect that leads to higher efforts by earlier-movers, lower efforts by later-movers, and higher total effort. These effects have different implications to different objectives a contest designer may have. 
The optimal contests are summarized by \cref{T:summary}. Perhaps the most surprising finding of the paper is that for most objectives the optimal contest is one of the three standard contests the previous literature has been focused on---the simultaneous contest, the sequential contest, or the first-mover contest.

\begin{table}[ht!]
    \centering  
    \begin{tabular}{cc|c|c}
       \multicolumn{2}{c|}{Objective function} & Minimizer & Maximizer \\
        \hline
        Total effort  & $X^*=\sum_{i=1}^n x_i^*$ & Simultaneous & Sequential \\ 
        \hline
        Total welfare & $\sum_{i=1}^n u_i(\bx^*)$ & \multirow{3}{*}{Sequential} & \multirow{3}{*}{Simultaneous} \\
        Lowest effort & $\lowx$ & & \\
        Lowest payoff & $\lowu$ & & \\
        \hline
        Highest effort& $\highx$ & Simultaneous & Single-leader$^{\dagger}$ \\
        Highest payoff& $\highu$ &  $(2,1,\dots,1)^{\dagger}$ & First-mover$^{\dagger}$ \\ 
        \hline
        Effort inequality & $\highx-\lowx$ & \multirow{2}{*}{Simultaneous} & Sequential$^{\dagger}$ \\
        Payoff inequality & $\highu-\lowu$ & & First-mover$^{\dagger}$ \\ 
    \end{tabular}
    \caption{Summary of the optimal contests for the $8 \times 2=16$ objectives discussed in the paper. 
    Contests marked with $^{\dagger}$ are optimal at least for (1) Tullock contests with $n\leq 12$, (2) contests with linear $h(X)$ function for any $n$, and (3) large contests.
    The simultaneous contest is $(n)$, the sequential contest is $(1,1,\dots,1)$, the first-mover contest is $(1,n-1)$, and single-leader contest are $(1,n_2,\dots,n_T)$ for all $n_2,\dots,n_T$ (including the sequential and the first-mover contests).
\label{T:summary}
    }
\end{table}

\clearpage

\bibliography{zotero_contest_architecture}
\bibliographystyle{ecta}

\newpage
\renewcommand{\appendixpagename}{Appendices for Online Publication}
\appendixpage
\appendix

\section{Notation, Assumptions, and Useful Results} \label{A:assumptions}

In this appendix, I describe additional notation and formalize technical assumptions that are necessary for formal proofs. I then describe a few useful results from the literature.

\subsection{Notation}

Instead of working directly with the marginal benefit function $h(X)$ it is more convenient to work with the following functions $g,g_1,\dots,g_T$. Each of these function captures a particular curvature property of $h$ function and can be interpreted as a higher-order strategic substitutability term.
\begin{definition}[Functions $g$ and $g_k$]
  Let $g(X) = -\frac{h(X)}{h'(X)}$. 
  With this, let us define $g_1,\dots,g_T$ recursively as $g_1(X)=g(X)$ and $g_{k+1}(X)=-g_k'(X) g(X)$.
\end{definition}
For example, in the case of linear $h(X) = a (\oX-X)$, $g(X)= \oX-X= g_k(X)$ for all $k$. In the case of normalized Tullock payoffs $u_i(\bx) = \frac{x_i}{X} - x_i$ and so $h(X)=\frac{1}{X}-1$ and $g(X)= X (1-X)$. Therefore $g_k(X)$ is a polynomial of degree $k+1$.

For a contest $\bn$ let us define its measures of information. Intuitively, it is a vector of integers, that captures the number of direct and indirect observations of other players' efforts. 
Let me describe the construction with an example of contest $\bn = (1,2,3)$. Each player observes its own effort. This is captured by the first measure $S_1(\bn)=1+2+3=6$. The second measure is the number of direct observations of other players' efforts, $S_2(\bn)=1\cdot 2 + 1 \cdot 3 + 2 \cdot 3=11$. The third measure captures observations of observations, i.e.\ influences through one connection. In particular, $S_3(\bn)=1 \cdot 2 \cdot 3 = 6$ (for example, player $1$ influences player $2$, who in turn influences player $4$). Therefore $\bS(\bn)=(6,11,6)$ in the example. 
\begin{definition}[Measures of Information $\bS(\bn)$]
  For a given contest $\bn=(n_1,\dots,n_T)$, its measures of information are defined as $\bS(\bn) = (S_1(\bn),\dots,S_T(\bn))$, where $S_k(\bn)$ is the sum of products of all possible $k$-combinations of the set $\{n_1,\dots,n_T\}$.
\end{definition}

For each period $t$, let $\bn^t$ denote the \emph{subcontest} starting after period $t$, i.e.\ $\bn^t = (n_{t+1},\dots,n_T)$. Then $\bS(\bn^t)$ are denoted analogously to the original contest.
For example, if $\bn=(1,2,3)$, then $\bn^1 = (2,3)$ and $\bn^2=(3)$, therefore $\bS(\bn^1)=(5,6)$ and $\bS(\bn^2)=(3)$.

Finally, let me define functions $f_0,\dots,f_T$ as follows\footnote{Note that when $t=0$, then $\bn^t = \bn$ is the whole contest.}
\begin{equation} \label{E:ftdef}
  f_t(X) = X - \sum_{k=1}^{T-t} S_k(\bn^t) g_k(X).
\end{equation}
These functions take the role of \emph{inverted best-response functions} that I discuss below. Intuitively, the inverse functions $f_t^{-1}(X_t)$ describe what is the total effort $X$ when after period $t$ the cumulative effort is $X_t$ and all players after period $t$ behave optimally.

\subsection{Technical Assumptions}

The first technical assumption is sufficient to guarantee that the best-response functions of players are well-behaved. This is a sufficient assumption to guarantee both the existence and uniqueness of the subgame-perfect Nash equilibrium for contest $\bn$.
\begin{assumption}[Inverted Best-Responses are Well-Behaved] \label{A:ass1}
  For all $t=0,\dots,T-1$, the function $f_t$ defined by \cref{E:ftdef} has the following properties:
  \begin{enumerate}
      \item $f_t(X)=0$ has a root in $[\uX_{t+1},\oX]$.\footnote{Function $f_T(X)=X$, therefore it has exactly one root, $\uX_T=0$.} Let $\uX_t$ be the highest such root.
      \item $f_t(X) < 0$ for all $X \in [\uX_{t+1},\uX_t)$.
      \item $f_t'(X)>0$ for all $X \in [\uX_t,\oX]$.
  \end{enumerate}
  Moreover, $\uX_0 \in (0,\oX)$.
\end{assumption}

The second assumption states that the efforts are higher-order strategic substitutes near equilibrium. This assumption is stronger than the standard assumption of strategic substitutes, which requires that higher effort of an opponent reduces the incentive for a player to exert effort. Higher-order influences matter in sequential games, where the earlier-mover may not only influence a follower directly but indirectly through changing the behavior of some players between them.
 
\begin{assumption}[Higher-Order Strategic Substitutes] \label{A:ass2}
  $g_k(X^*) > 0$ for all $k=2,\dots,T$.
\end{assumption}

\cite{hinnosaar_optimal_2018} shows that the two assumptions are satisfied for Tullock contests with at least three players as well as in many other situations with well-behaving marginal benefit functions.

\subsection{Useful Results}

The proofs rely on a few useful results from earlier works. The main tool for the analysis is the characterization theorem, which characterizes the unique equilibrium for any contests.

\begin{theorem}[Characterization Theorem \citep{hinnosaar_optimal_2018}] \label{T:characterization}
  Under \cref{A:ass1} the total equilibrium effort $X^*$ is the highest root of $f_0(X)=0$, where
  \begin{equation}
    f_0(X)
    = X - \sum_{k=1}^T S_k(\bn) g_k(X).
  \end{equation}
  The equilibrium effort of player $i$ in period $t$ is $x_i^* = g_1(X^*) + \sum_{k=1}^{T-t} S_k(\bn^t) g_{k+1}(X^*)$.
\end{theorem}

\paragraph{Informativeness and equilibria:}
As \cite{hinnosaar_optimal_2018} shows, this result has strong implications for equilibrium behavior. The total equilibrium effort $X^*$ increases with the informativeness of the contest. Therefore it is minimized by the simultaneous contest (the least informative contest) and maximized by the sequential contest (the most informative contest).

\paragraph{Independence of permutations:} permutations in $\bn$ do not change the total equilibrium effort $X^*$. For example, the contest $\bn=(1,2,3)$ has exactly the same total effort as the contest $\hbn=(1,3,2)$ because $\bS(\hbn)=(6,11,6)=\bS(\bn)$. Of course, individual efforts may be affected, as they also depend on subcontest. But note that permutations within the subcontest do not affect individual equilibrium effort either: in the examples here, $\bn^1 = (2,3)$ and $\hbn^1=(3,2)$ is a permutation. Therefore $\hx_1^* = x_1^*$.
    
\paragraph{Earlier-mover advantage:} take two players, $i$ from period $t$ and $j$ from a later period $s>t$. Then $x_i^* > x_j^*$ and $u_i(\bx^*) > u_j(\bx^*)$.
As explained in the text, this means that highest effort is always $\highx = x_1^*$, lowest effort $\lowx = x_n^*$, highest payoff $\highu = u_1(\bx^*)$, and lowest playoff $\lowu = u_n(\bx^*)$.

\paragraph{Large contests:}
Moreover, \cite{hinnosaar_stackelberg_2019} shows that in the limit when the number of players becomes large, the equilibrium behavior converges to specific functional form. Intuitively, as the number of players becomes large and the total equilibrium effort increases with the number of players, it is quite natural that the total equilibrium effort converges to its full dissipation limit $\oX$. As the function $h(X)$ is smooth, we can approximate it with a linear function near $\oX$ arbitrarily closely when the number of players is large. Therefore it is not surprising that the equilibrium behavior of individual players converges to equilibrium behavior of the same game but with linear $h(X)$ function. 

\begin{theorem}[Competitive Limits \citep{hinnosaar_stackelberg_2019}] \label{T:limits}
  Fix a sequence $\bn = (n_1,\dots,n_T)$ and let us increase $n_t$ at a particular period $t$, while keeping other elements of the sequence $\bn$ unchanged. Then $\lim_{n_t \to \infty} X^* = \oX$ and for each player $i$ arriving in period $s$,
  \begin{equation} \label{E:limit}
    \lim_{n_t \to \infty} x_i^*
    = \begin{cases}
        0 & \forall s \geq t, \\
        \frac{\oX}{\prod_{k=1}^s (1+n_k)} & \forall s < t.
      \end{cases}
  \end{equation}
\end{theorem}

I use this result as follows. Note that $n_t$ is not determined, as with different $n$ the contest designer may want to use different disclosures. I use \cref{E:limit} as an approximation for the equilibrium behavior, i.e.\ each player $i \in \prt_s$ chooses
\begin{equation} \label{E:approxxi}
  x_i^* \approx \frac{\oX}{\prod_{k=1}^s (1+n_k)}.
\end{equation}
Note that this approximation also implies that
\begin{equation}
  X^* = \sum_{i=1}^n x_i^* \approx \oX - \frac{\oX}{\prod_{k=1}^T (1+n_k)}.
\end{equation}
Since $h(X)$ is a smooth function and $X^* \to \oX$, we can approximate $h(X^*)$ linearly near $\oX$, i.e.\
\begin{equation} \label{E:approxh}
  h(X^*) 
  \approx h(\oX) + h'(\oX) (\oX-X^*)
  = \frac{\alpha \oX}{\prod_{k=1}^T (1+n_k)}, \text{ where }\alpha = -h'(\oX) > 0.
\end{equation}

\section{Proofs} \label{A:proofs}

\subsection{Proof of \texorpdfstring{\Cref{P:totals}}{Proposition 1} (Total Effort and Total Welfare)}

Results for $X^*$ follow from \cref{T:characterization}. The total welfare is $W(\bx^*)=X^* h(X^*)$, so
\begin{equation}
  \dd{W(\bx^*)}{X^*}
  = h(X^*) + X^* h'(X^*)
  \geq 0 
  \;\;\iff\;\;
  X^* \geq - \frac{h(X^*)}{h'(X^*)} = g(X^*).
\end{equation}
Note that by \cref{T:characterization}, each $x_i^* \geq g(X^*)$, therefore this condition is always satisfied. Each disclosure increases $X^*$ and therefore decreases total welfare.

\subsection{Proof of \texorpdfstring{\Cref{P:lowest}}{Proposition 2} (Lowest Effort and Lowest Payoff)}

As argued in the text, $\lowx = x_n^*$ and by \cref{T:characterization}, 
  $x_n^* 
  = g(X^*)$.
Since $g(X^*)$ is strictly decreasing in $X^*$, the lowest effort $x_i^*$ is minimized when $X^*$ is maximized, i.e.\ by sequential contest, and maximized when $X^*$ is minimized, i.e.\ by simultaneous contest.
Similarly, $\lowu = u_n(\bx^*) = x_n^* h(X^*) = g(X^*) h(X^*)$, where both $g$ and $h$ are decreasing in $X^*$, which leads to the same conclusion.

\subsection{Proof of \texorpdfstring{\Cref{P:highestx}}{Proposition 3} (Highest Effort)}

Let us first consider minimization of the highest equilibrium effort $\highx  = x_1^*$. Let $X^{sim}$ be the total equilibrium effort from simultaneous $n$-player contest $(n)$. Then clearly the highest equilibrium effort $x_i^{sim} = \frac{X^{sim}}{n}$. Now take any non-simultaneous contest. As this contest is strictly more informative than the simultaneous contest, $X^* > X^{sim}$. Moreover, by the earlier-mover advantage result, the highest effort is strictly larger than the average, which proves the claim, since
\begin{equation}
  x_1^* > \frac{X^*}{n} > \frac{X^{sim}}{n} = x_1^{sim}.
\end{equation}

Suppose now that $\bn$ maximizes the highest effort, which by \cref{T:characterization} is
\begin{equation}
  \highx = x_1^* = g_1(X^*) + \sum_{k=1}^{T-1} S_k(\bn^1) g_{k+1}(X^*).
\end{equation}
I first claim that $n_1 < n_t$ for all $t$. If this is not the case, we can take a permutation of $\bn$, which leaves $X^*$ unchanged, but increases the information measures of the subcontest $\bn^1$, therefore increasing the highest effort. 
Next, I claim that $n_1 = 1$. Suppose that this is not the case, i.e.\ $2 \leq n_1$ But the previous observation, $n_2 \geq n_1 \geq 2$. Consider an alternative contest $\hbn = (n_1-1, n_2+1,n_3,\dots,n_T)$. Then $\hbn^1 = (n_2+1,n_3,\dots,n_T)$. Since $n_2 \geq n_1$, we have 
\[
  (n_1 -1)(n_2+1) = n_1 n_2 - n_2 + n_1 -1 < n_1 n_2.
\]
Therefore $S_2(\hbn) < S_2(\bn)$ and $\bS(\hbn) < \bS(\bn)$. Therefore $\hX^* < X^*$. This means that each $g_k(\hX^*) > g_k(X^*)$. Moreover, clearly $\bS(\hbn^1) > \bS(\bn^1)$. Therefore $\hx_1^* > x_1^*$, which means that $x_1^*$ was not the maximal highest effort.

Finally, when the $h(X)$ is linear or when number of players is large, then by the arguments above, the highest effort is (approximately) $x_1^* \approx \frac{\oX}{1+n_1}$. Clearly, this is maximized by setting $n_1 = 1$, i.e.\ a disclosure right after the first player. All other disclosures have no impact when $h(X)$ is linear and a negligible impact when $n$ is large enough.

\subsection{Proof of \texorpdfstring{\Cref{P:highestu}}{Proposition 4} (Highest Payoff)}

Let $\bn$ be a contest that minimizes the highest equilibrium payoff $\highu = u_1(\bx^*) = x_1^* h(X^*)$. 
I claim that $n_1 \geq n_t$ for all $t$. If this is not the case, then there exists a permutation of $\bn$ such that $x_1^*$ is increased and $X^*$ and thus $h(X^*)$ is unchanged. This would violate the optimality of $\bn$.
When $n$ is large, then by the highest equilibrium payoff is approximately
\begin{equation}
  u_1(\bx^*)
  = x_1^* h(X^*)
  \approx \frac{\oX}{1+n_1} \frac{\alpha \oX}{\prod_{s=1}^T (1+n_s)}.
\end{equation}
Minimizing this objective is equivalent to maximizing $(1+n_1)^2 \prod_{s=2}^T (1+n_s)$. We can do it in two steps. First, fix $n_1 \geq 1$ and choose optimal subcontest, which therefore needs to maximize $\prod_{s=2}^T (1+n_s)$ subject to constraint that $\sum_{s=1}^T n_s = n-n_1$. Splitting each $n_s \geq 2$ onto $n_s' >0$ and $n_s''>0$ always increases the product as $(1+n_s')(1+n_s'')=1+n_s'+n_s''+n_s'n_s''>1+n_s'+n_s''$. Therefore the maximized product $\prod_{s=2}^T (1+n_s) = 2^{n-n_1}$ and so the maximization problem is
\begin{equation}
  2^n \max_{1 \leq n_1 \leq n} (1+n_1)^2 2^{-n_1}.
\end{equation}
Treating $n_1$ as a continuous variable and differentiating the objective gives
\begin{equation}
  -2^{-n_1} (1 + n_1) \left(
    (1+n_1) \log 2 - 2
  \right)
  \leq 0
  \;\;\;
  \iff
  \;\;\;
  n_1 \leq \frac{2}{\log 2} - 1 \approx 1.8854.
\end{equation}
The objective is decreasing in $n_1$ for all $n_1 \geq 2$, so there are only two candidates for the maximizer: either $n_1=1$ or $n_1=2$. Direct comparison gives 
\begin{equation}
  (1+1)^2 2^{-1}
  = 2
  <
  (1+2)^2 2^{-2}
  = \frac{9}{4}.
\end{equation}
Therefore the optimal large contest is $\bn=(2,1,\dots,1)$.

Let $\bn$ be a maximizer of the highest payoff $u_1(\bx^*) = x_1^* h(X^*)$. The same arguments as in the proof of \cref{P:highestx} show that $n_1$ must be equal to $1$ (first $n_1 \leq n_t $ for all $t$, otherwise can take a permutation, and then if $n_1 > 1$, splitting it makes the contest less homogeneous and thus decreases $X^*$, which increases total payoff while also increasing player $1$'s share in it). 
Now, let $\hbn$ be a contest that maximizes the highest effort. Let the corresponding effort profile be $\hbx^*$. Then by definition $x_1^* \leq \hbx^*$ and $u_1(\bx^*) = x_1^* h(X^*) \geq \hx_1^* h(\hX^*) = u_1(\hbx^*)$, which cannot be satisfied unless $X^* \leq \hX^*$. Therefore $\bn$ cannot be more informative than $\hbn$.

The final claim follows from the arguments above. With linear $h(X)$ or large $n$, one contest that maximizes the highest effort is $\hbn = (1,n-1)$. As the highest payoff maximizer must have $n_1=1$ and cannot be more informative than $\hbn$, it must coincide with $\hbn$.

\subsection{Proof of \texorpdfstring{\Cref{P:inequality}}{Proposition 5} (Effort Inequality and Payoff Inequality)}

As argued in the text, the simultaneous contest is the only contest where the efforts and payoffs of all players are equal, so it is the unique minimizer of both the equilibrium effort inequality $\highx-\lowx = x_1^* - x_n^*$ as well as the equilibrium payoff inequality $\highu-\lowu = u_1(\bx^*) - u_n(\bx^*)$. 

By the same arguments as with the highest effort and with the highest payoff, we get that $n_1=1$ both with effort inequality and payoff inequality maximizers. Note that since $x_1^*$ is independent of permutations of the subcontest $\bn^1$ and $x_n^* = g(X^*)$ is independent of permutations in the whole contest, both inequalities are independent of permutations in $\bn^1$.

Suppose that $\bn$ is an effort inequality maximizer and let the corresponding equilibrium effort profile be $\bx^*$. Take any highest effort maximizer $\hbn$ and let the corresponding effort profile be $\hbx$. Then by construction, $x_1^* \leq \hx_1^*$ and $x_1^*-x_n^* \geq \hx_1^* -\hx_n^*$, which implies that $g(X^*) = x_n^* \leq \hx_n^* = g(\hX^*)$. As $g(X^*)$ is decreasing in $X^*$ it implies $X^* \geq \hX^*$, which means that $\bn$ cannot be less informative than $\hbn$.
Finally, if $h(X)$ is linear or $n$ is large, then any single-leader contest maximizes the highest effort. One such contest is the sequential contest, which is more informative than any other contest. Therefore it must be the unique maximizer of the effort inequality in these cases.

Now, let $\bn$ be the maximizer of the equilibrium payoff inequality. As argued above, $n_1=1$.
Take any maximizer of the highest payoff $\hbn$. Then 
$u_1(\bx^*) - u_n(\bx^*)\geq u_1(\hbx^*) - u_n(\hbx^*) \geq  u_1(\bx^*) - u_n(\hbx^*)$ and therefore $g(X^*) h(X^*) = u_n(\bx^*) \leq u_n(\hbx^*) = g(\hX^*) h(\hX^*)$, which implies that $X^* \geq \hX^*$. This means that $\bn$ cannot be less informative than $\hbn$.
Next, suppose that $\hbx$ maximizes the effort inequality. Then $(x_1^*-x_n^*) h(X^*) \geq (\hx_1^*-\hx_n^*) h(\hX^*) \geq (x_1^*-x_n^*) h(\hX^*)$, which implies $h(X^*) \geq h(\hX^*)$ or equivalently $X^* \leq \hX^*$. Therefore $\bn$ cannot be more informative than $\hbn$.

When $h(X)$ is linear or $n$ is large, the results above do not tell us much, since none of the single-leader contests is less informative than $(1,n-1)$ (the maximizer of the highest payoff) and no contest is more informative than the sequential contest (the maximizer of the effort inequality). Therefore we have to proceed with direct proof. The equilibrium payoff inequality is (approximately)
\begin{align}
    u_1(\bx^*)-u_n(\bx^*)
    &\approx \frac{\oX}{1+n_1} \frac{\alpha \oX}{\prod_{s=1}^T (1+n_s)} - \frac{\oX}{\prod_{s=1}^T (1+n_s)} \frac{\alpha \oX}{\prod_{s=1}^T (1+n_s)} 
    \notag \\
    &= \frac{\alpha \oX^2}{(1+n_1)^2} 
    \frac{1}{\prod_{s=2}^T (1+n_s)}
    \left[
      1 - \frac{1}{\prod_{s=2}^T (1+n_s)}
    \right].
\end{align}
I already proved that $n_1=1$ for any payoffs, including linear. The remaining question is how to arrange the remaining $n-1$ players into subcontest $\bn^1=(n_1,\dots,n_T)$. Let us denote $Y(\bn^1)=\frac{1}{\prod_{s=2}^T (1+n_s)}$ for brevity. Then $Y(\bn^1)$ is decreasing with each disclosure. Combining this with the fact that the optimal contest is not simultaneous, we get $Y(\bn^1) \leq \frac{1}{1+n-n_1} \leq \frac{1}{2}$. Now, note that the expression $Y(\bn^1) [1-Y(\bn^1)]$ is strictly increasing in $Y(\bn^1)$ for all $Y(\bn^1) < \frac{1}{2}$. Therefore maximization of payoff inequality requires that there are no disclosures after period $1$, i.e.\ $Y(\bn^1) = \frac{1}{1+n-n_1}=\frac{1}{n}$. 
This implies that the optimal contest is the first-mover contest $\bn=(1,n-1)$.

\clearpage
\section{Additional Figures} \label{A:figures}

\begin{figure}[!ht]%
    \centering
    \subfloat[Total effort]{\includegraphics[trim={0.22cm 0.35cm 0.86cm 0.35cm},clip,width=0.45\linewidth]{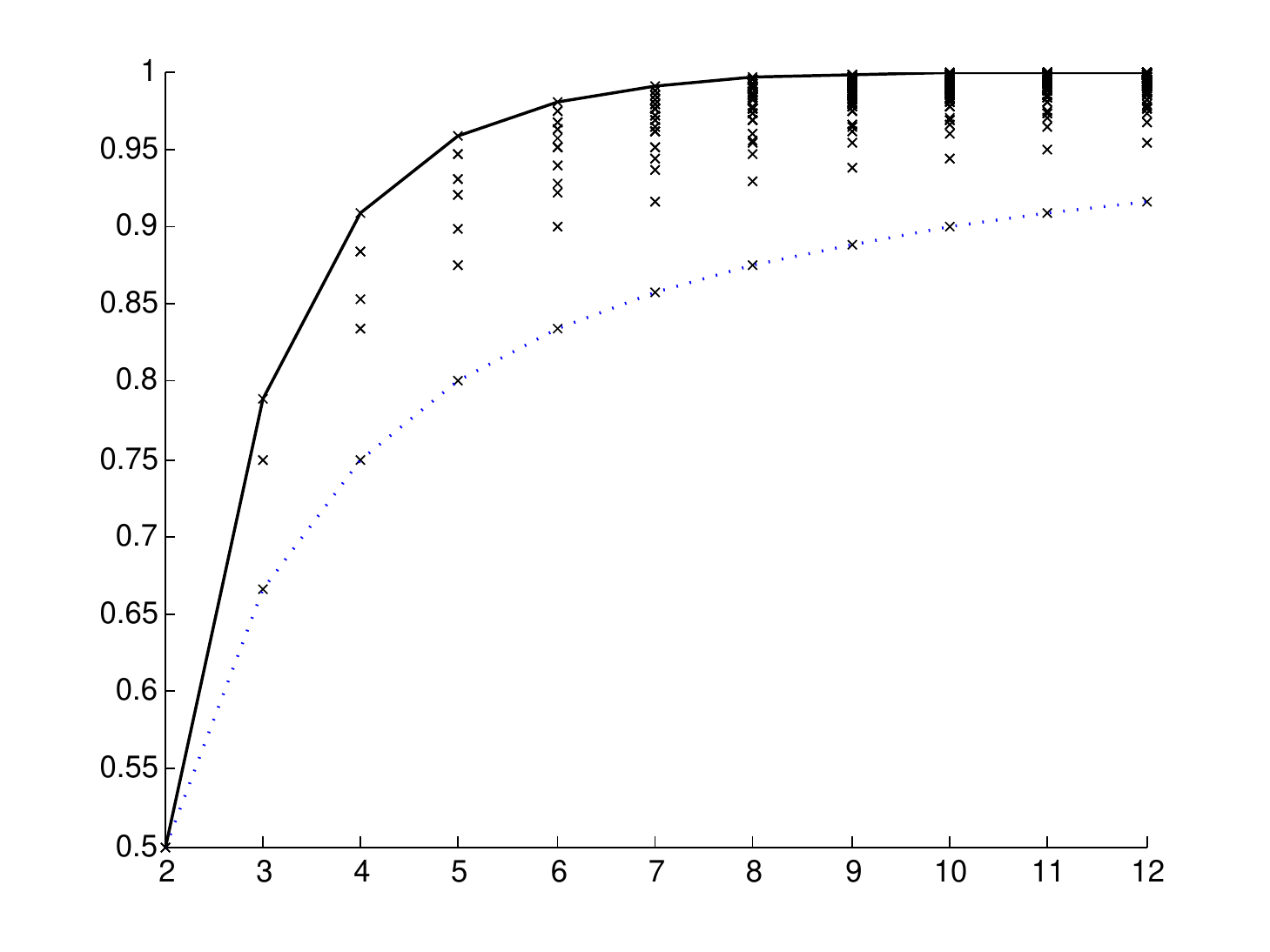} \label{F:plotX}}
    \qquad
    \subfloat[Total welfare]{\includegraphics[trim={0.22cm 0.35cm 0.86cm 0.35cm},clip,width=0.45\linewidth]{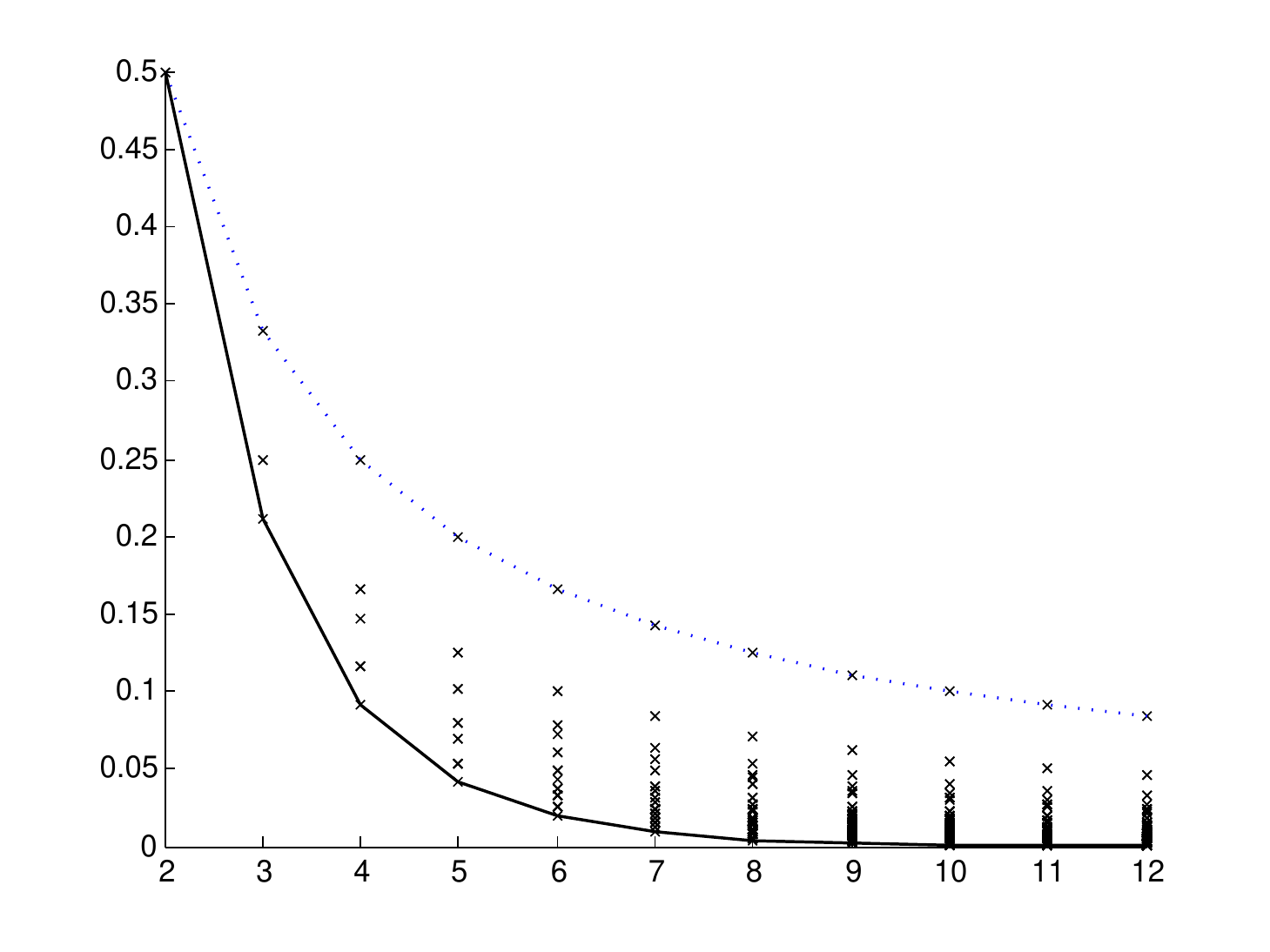} \label{F:plotW}}
    \caption{Total equilibrium effort $X^*$ and total equilibrium welfare $W(\bx^*)$ in all normalized Tullock contests with $2 \leq n \leq 12$ players. The solid black line indicates sequential contests and dashed blue line simultaneous contests.
    \label{F:plotXW}}%
\end{figure}

\begin{figure}[!ht]%
    \centering
    \subfloat[Lowest effort]{\includegraphics[trim={0.22cm 0.35cm 0.86cm 0.35cm},clip,width=0.45\linewidth]{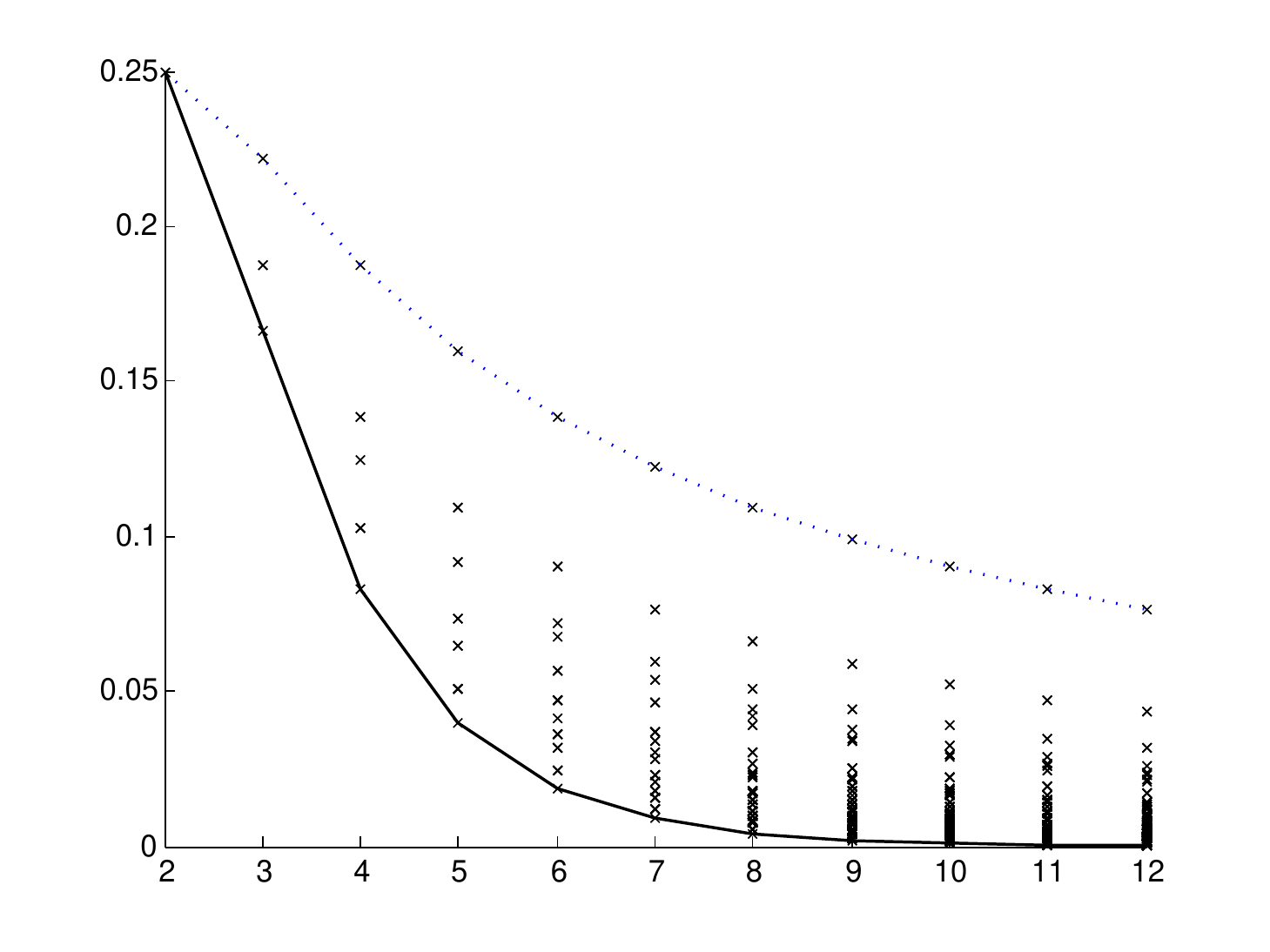} \label{F:plotxn}}
    \qquad
    \subfloat[Lowest payoff]{\includegraphics[trim={0.22cm 0.35cm 0.86cm 0.35cm},clip,width=0.45\linewidth]{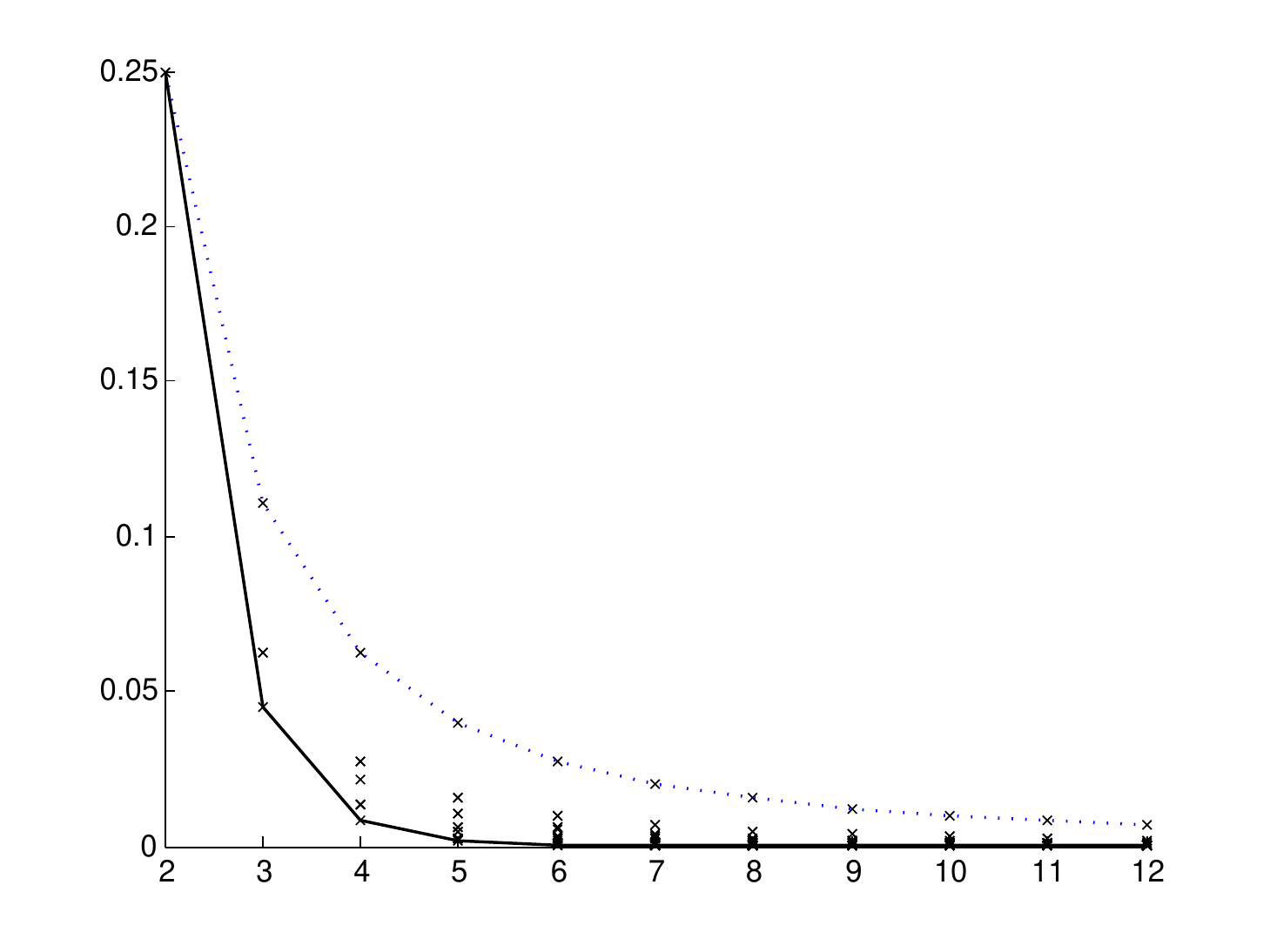} \label{F:plotun}}
    \caption{Lowest effort $\lowx$ and lowest payoff $\lowu$ in all normalized Tullock contests with $2 \leq n \leq 12$ players. The solid black line indicates sequential contests and dashed blue line simultaneous contests.
    \label{F:plotxnun}}%
\end{figure}

For completeness, I include here the figures describing the total equilibrium effort (\cref{F:plotX}), total equilibrium welfare (\cref{F:plotW}), lowest equilibrium effort (\cref{F:plotxn}), and lowest equilibrium payoff (\cref{F:plotun}) in all Tullock contests with $2$ to $12$ players. The figures confirm fully general findings discussed in the text---in each of these eight possible cases the optimal contest is either the sequential contest or the simultaneous contest.

\clearpage

\section{Minimizing or Maximizing the \texorpdfstring{$i$}{i}-th Highest Effort and Payoff} \label{A:ithlargest}

In the case of linear $h(X)$ or large $n$, the problem is simple enough to consider other objectives. Consider for example $i$-th highest effort. By the earlier-mover advantage result, it is the effort of player $i \in \prt_t$. If $h(X)$ is linear or $n$ is large, it is (approximately)
\begin{equation}
  x_i^*
  \approx 
  \frac{\oX}{\prod_{k=1}^t (1+n_k)}.
\end{equation}
Remember the general trade-off: each disclosure reduces the marginal benefit of effort for all players and therefore pushes all efforts downwards, but it increases the efforts of earlier-movers whose effort is now made public. Therefore each disclosure before the arrival of player $i$ only has the former effect and therefore reduces $x_i^*$. Therefore minimizing $x_i^*$ requires players before $i$ to be arranged sequentially and maximizing requires them to be simultaneous (i.e.\ no disclosures prior to $i$). 
Disclosures after player $i$ have the two opposite effects, so their impact depends on the relative magnitudes. However, in the case considered here (linear demand), they are exactly equal and therefore balance out. Therefore $x_i^*$ is independent on the choice of disclosures after the period player $i$ arrives.\footnote{This is the \emph{Stackelberg Independence} property of a standard sequential homogeneous goods oligopoly, studied in detail in \cite{hinnosaar_stackelberg_2019}.}
This allows us to state the following simple proposition.
\begin{proposition}
  When $h(X)$ is linear or $n$ is large
  \begin{enumerate}
      \item The contest that has $i-1$ first players arranged sequentially and the rest of the players arriving simultaneously, i.e.\ $\bn=(1,\dots,1,n+1-i)$, minimizes the $i$-th largest effort.
      \item Any contest that has $i$ players in the first period, i.e.\ $\bn=(i,n_2,\dots,n_T)$, maximizes the $i$-th largest effort.
  \end{enumerate}
\end{proposition}
When $i=n$ and $i=1$ this result confirms \cref{P:lowest,P:highestx}: lowest effort ($i=n$) is minimized by sequential and maximized by sequential contest, and highest effort ($i=1$) is minimized by the simultaneous contest and maximized by any single-leader contest.


We can analogously study the $i$-th highest payoff $u_i(\bx^*)$, which is (approximately) 
\begin{equation}
  u_i(\bx^*) 
  = x_i^* h(X^*)
  \approx
  \frac{\oX}{\prod_{k=1}^t (1+n_k)} \frac{\alpha \oX}{\prod_{k=1}^T (1+n_k)}.
\end{equation}
Now there are three effects. First, all disclosures before the arrival of player $i$ decrease both $x_i^*$ and $h(X^*)$ and thus decrease $u_i(\bx^*)$. Second, all disclosures after period $t$ (when $i$ arrives) leave $x_i^*$ unaffected, but decrease $h(X^*)$, thus decreasing $u_i(\bx^*)$. And third, the length of period $t$, i.e.\ delaying the first disclosure after the arrival of player $i$, has two opposite effects---it reduces $x_i^*$ but increases $h(X^*)$.
Combining these effects gives us the following proposition.
\begin{proposition}
  When $h(X)$ is linear or $n$ is large
  \begin{enumerate}
      \item The contest that minimizes the $i$-th highest payoff has the following structure: $i-1$ sequential players first, then two players simultaneously, and then the remaining $n-i-1$ players sequentially.
      \item The contest that maximizes the $i$-th highest payoff is $\bn=(i,n-i)$ when $i \leq \overline{i}=-\frac{1}{2} + \sqrt{\frac{5}{4}+n}$ and the simultaneous contest when $i \geq \overline{i}$.
  \end{enumerate}
\end{proposition}
This confirms the findings from \cref{P:lowest} (the lowest payoff is minimized by sequential and maximized by simultaneous contest) and \cref{P:highestu} (the highest payoff is minimized by contest $(2,1,\dots,1)$ and maximized by the first-mover contest).
\begin{proof}
Let us first consider minimization of the $i$-th highest payoff $u_i(\bx^*)$.
The arguments state that the players before $i$ should be sequential as well as players after period $t$. The remaining question is $n_t$, i.e.\ how many player to pool with $i$. This gives a maximization problem
\begin{equation}
  \min_{1 \leq n_t \leq n+1-i} 
  \frac{1}{2^{i-1} (1+n_t)} \frac{1}{(1+n_t) 2^{n-n_t}}
  \;\;\iff\;\;
  \max_{1 \leq n_t \leq n+1-i} (1+n_t)^2 2^{-n_t}.
\end{equation}
Treating $n_t$ as a continuous variable and differentiating the objective gives necessary condition for optimality
\begin{equation}
  2^{-n_t} (1+n_t) \left[ 2 - (1+n_t) \log 2 \right] = 0.
\end{equation}
Solving this gives $n_t = \frac{2}{\log 2}-1 \approx 1.8854$. It is straightforward to verify that the objective is concave for all $1 \leq n_t \leq n$, so the only two candidates for the optimum are $n_1=1$ and $n_1=2$. The latter is bigger than the former as $(1+2)^2 2^{-2} = \frac{9}{4} > (1+1)^2 2^{-1}= 2$. Therefore optimal $n_t=2$. We get that the optimal contest is $\bn = (\underbrace{1,\dots,1}_{i-1},2,\underbrace{1,\dots,1}_{n-i-1})$.

Now let us consider maximization of the $i$-th highest payoff $u_i(\bx^*)$. As argued above, all players before $i$ should be pooled with $i$, so that player $i$ arrives in period $1$. All players after period $t$ should be pooled together as well, which would lead to a two-period contest $(n_1,n-n_1)$ with $i \leq n_1 \leq n$ (in case $n_1=n$ it is in fact a simultaneous contest). Therefore the maximization problem is equivalent to
\begin{equation}
  \max_{i \leq n_1 \leq n}
  \frac{1}{1+n_1} \frac{1}{(1+n_1)(1+n-n_1)}
  \;\; \iff \;\;
  \min_{i \leq n_1 \leq n}
  (1+n_1)^2 (1+n-n_1).
\end{equation}
This problem is concave in the interior, therefore one of the two corners is optimal, either 
$n_1=i$, which gives $(1+i)^2 (1+n-i)$ as the value of the objective, or $n_1=n$ which gives $(1+n)^2$. The former is the minimizer if and only if 
\begin{equation}
  (1+i)^2 (1+n-i) \leq (1+n)^2.
\end{equation}
Equalizing left-hand-side with the right-hand-side gives and solving for $i$ gives three roots $n$ and $-\frac{1}{2} \pm \sqrt{\frac{5}{4}+n}$. Lowest root is always negative, and the second one $\overline{i} = -\frac{1}{2} + \sqrt{\frac{5}{4}+n}$ always strictly between $1$ and $n$.\footnote{Since 
$1 < \overline{i} = -\frac{1}{2} + \sqrt{\frac{5}{4}+n} \iff 1 < n$ and
$n > \overline{i} 
= -\frac{1}{2} + \sqrt{\frac{5}{4}+n} 
\iff 
n^2 > 1$.}
Moreover, at $i=0$ the condition is clearly violated. Therefore, we get that for low values $i \leq \overline{i}$ the condition is violated and therefore the optimal contest is $\bn=(i,n-i)$, whereas for high values $i \geq \overline{i}$ the condition is satisfied and therefore the optimal contest is simultaneous $\bn=(n)$.
\end{proof}

\end{document}